\documentstyle[preprint,aps,psfig]{revtex}

\def\cal{\rm}
\def\be{\begin{equation}}
\def\ee{\end{equation}}
\def\bea{\begin{eqnarray}}
\def\eea{\end{eqnarray}}

\def\Journal#1#2#3#4{{#1} {\bf #2}, #3 (#4)}

\def\NPA{{\em Nucl. Phys.} A}
\def\PLB{{\em Phys. Lett.}  B}
\def\PR{{\em Phys. Rep.} }
\def\PRL{\em Phys. Rev. Lett.}
\def\PRD{{\em Phys. Rev.} D}
\def\PRB{{\em Phys. Rev.} B}
\def\ZPC{{\em Z. Phys.} C}

\begin{document}
\draft
\preprint{}
\title{Phase transition of a finite quark-gluon plasma
\footnote[1]{Supported by GSI, BMBF, DFG}}
\author{C.~Spieles, H.~St\"ocker}
\address{Institut f\"ur
Theoretische Physik,  J.~W.~Goethe-Universit\"at,\\
D-60054 Frankfurt am Main, Germany}
\author{C.~Greiner}
\address{Institut f\"ur Theoretische Physik,
J.~Liebig-Universit\"at,\\
D-35392 Gie{\ss}en, Germany}
\date{\today}
\maketitle
\begin{abstract}
The deconfinement transition region between hadronic matter and quark-gluon
plasma is studied for finite volumes. Assuming simple model equations of state
and a first order phase transition, we find that fluctuations in finite
volumes hinder a sharp
separation between the two phases around the critical temperature, leading
to a {\em rounding} of the phase transition. 
For reaction volumes expected
in heavy ion experiments, the softening of the equation of state is reduced
considerably. This is especially true when the requirement of exact color-singletness is
included in the QGP equation of state.
\end{abstract}

\newpage
\section{Motivation}
A primary goal of relativistic nuclear collisions is the
observation of a phase transition of confined, hadronic matter to
a deconfined quark-gluon plasma. One of the proposed signatures, namely
hydrodynamic flow,  is based on
the presumed softening of the eqution of state due to the rapid increase of
the entropy density. It has been investigated in the framework of
relativistic hydrodynamic models \cite{Hun95,Ris96a}. 
The expansion of once compressed matter is predicted to
be delayed in the case of a QGP, which in turn leads to a reduction of the 
transverse (directed) flow \cite{Bra94,Ris95} 
\footnote{However, in a three-fluid 
hydrodynamical model \cite{Bra97} the directed nucleon flow is already
lowered as compared to the usual one-fluid models (which assume instantaneous
local thermalization between projectile and target). It will be exciting to
learn whether the softening of the EOS is also signaled in this model.}. 
This is mainly due to the fact
that the sound velocity vanishes for energy densities in the
mixed phase.
A smooth crossover transition within an assumed interval of 
$\Delta T=0.1 T_C$, on the 
other hand, results in drastically reduced
time delays as compared to a sharp transition \cite{Ris96b}. 

As is well known, a rounding of sharp first order phase transitions is
expected due to explicit finite size effects \cite{Bin84}.
The importance of fluctuations of the coexisting phases in small volumes 
of strongly interacting matter has
already been pointed out in \cite{Cse86} for the case of a liquid-gas phase
transition, and in \cite{Cse94} for the deconfinement phase transition. 
As a consequence, it was claimed
that the observation of two separate phases in heavy ion collisions might be 
hindered. In this work we explore this behaviour in more detail, 
starting from rather simple model equations of state. We put special emphasis on
the question, how the requirement of color singletness of the QGP phase
affects the phase transition on top of the fluctuation effect.

The limited reaction size of a heavy ion collision is a 
generally ignored problem of the experimental search for a quark-gluon plasma. 
According to one-fluid dynamical model calculations, one expects to produce 
deconfined matter in volumes of about $10-50\, \rm fm^3$ at fixed target 
energies \cite{Bra94}. Similar values of typical QGP cluster sizes are found
within a microscopic approach in combination with a percolation model
\cite{Wer94}. At collider energies the relevant reaction volumes for heavy
systems are 
estimated to be much larger: assuming an initial interaction 
volume of $V_0=\pi(1.15A^{1/3})^2 \tau_0$ with a formation 
time $\tau_0 \approx 1 \rm\,
fm/c$ and subsequent expansion one may expect (at the onset of confinement)
$V\approx 750\rm\, fm^3$ for Au+Au collisions at RHIC energies
($\sqrt{s}=200\,\rm AGeV$) and $V\approx 1350\rm\, fm^3$ for Pb+Pb at LHC
energies ($\sqrt{s}\approx 5.5\,\rm ATeV$)
 \cite{Sat92}.

However, the longitudinal flow velocities exceed the thermal motion by far. 
Thus, only subsystems of smaller volume can be
regarded as being in approximate ``global'' thermal equilibrium.
Only the latter are suitable
for the study of finite size effects. Therefore, if we require the local thermal motion
to be of the same order as the relative flow velocity,
one must restrict the study of finite size effects to regions of about one 
unit of rapidity (cf.~\cite{Spi96}). The volume of such (possibly multiple)
 subsystems would 
thus only be $V<100\rm \, fm^3$ for both RHIC and LHC.

Another motivation for the study of
finite size effects are lattice QCD calculations, which have to be
extrapolated from a finite lattice (although with periodic boundary
conditions) to infinite matter \cite{Lae96,Mey96}. Depending on the number 
of grid points in the spatial direction,
 the presently accesssible volumes are in the range of $V\approx 50-100\,\rm fm^3$.
The extrapolation to infinity is not clear \cite{Mey96}. Our
investigation can yield some insight what can be
phenomenologically expected when going from finite volumes to the infinite
volume limit.

\section{Phase coexistence and fluctuations}
Let us consider a finite volume of strongly interacting matter in
the grand canonical ensemble with vanishing net-baryon
density.  The model equation of state is 
constructed by matching a low energy density phase (a hadron gas) 
and a high energy density phase (the quark-gluon plasma):
For infinite matter the system undergoes a first order phase transition 
at a critical temperature $T_{\rm C}$. 
However, when we assume a finite volume $V$ of the system, 
statistical (thermodynamical) fluctuations are 
not negligible; they do not allow for a sharp transition between the 
two phases in this case. 
In general, the probability $p$ of finding the system in a 
state $\bf x$ is
given by $ p({\bf x}) \sim \exp{(-\beta F({\bf x}))} $, where $F({\bf x})$
is the free energy of the system \cite{Lan76,Cse94} (see also the appendix). 
Let us introduce an ``order'' parameter $\xi$ for the quantitative
characterization of the macroscopic state of the total system, such that
$\xi=1$ corresponds to the pure hadron phase and $\xi=0$
corresponds to the pure quark-gluon phase. For the mixed phase 
the fractional volumes are defined as
$V_{\rm h}=\xi V$
and $V_{\rm q}=(1-\xi) V$.

It is now assumed that the partition function of the total system factorizes 
into  the partition functions of the two individual phases
for fixed $\xi$ (see appendix). Then there are no correlations between the possible
microscopic energy eigenstates of the two subsystems (the two phases).
The free energy of the total system can thus be expressed in terms of the free
energy densities of the individual phases $f_{\rm q}$ and $f_{\rm h}$ and
the volume fraction of the hadron phase $\xi$:
\begin{equation}
\label{freeenergy}
F_\xi(T,V)=[f_{\rm h}(T,\xi V)\xi+f_{\rm q}(T,(1-\xi)V)(1-\xi)]V
\end{equation}
The normalized probability density is a function of the order parameter:
\begin{equation}
\label{probab}
p(\xi)=
\frac{\exp{(-\beta F_\xi(T,V))}}{\int_0^1{\rm d}\xi\exp{(-\beta F_\xi(T,V))}}
\end{equation}
The expectation value of any intensive thermodynamic quantity, $A(T,V)$, is then
calculated as:
\begin{equation}
\label{masterquant}
A(T,V)=\int_0^1{\rm d}\xi\, p(\xi;T,V) [A_h(T,\xi V)\,\xi+A_q(T,(1-\xi) V)(1-\xi)] \quad .
\end{equation}
A reminder of a thermodynamic derivation of (\ref{probab}) and (\ref{masterquant}) can be 
found in the appendix.

As long as explicit finite size effects in the equations of state of the 
individual phases are ignored (i.e. the free energies
$F_i$ are linear functions of $V_i$), the free energy
densities equal the negative pressure within the two phases:
\begin{equation} 
P_i(T)=-\frac{\partial F_i}{\partial V_i}=-\frac{F_i}{V_i}=-f_i
\end{equation}
When inserting this into (\ref{freeenergy}) and (\ref{probab}), 
respectively, one sees immediately that for large volumes, $V$, of the 
total system and a given
temperature, $T$, the occurence of the phase with the lower pressure is
strongly 
suppressed. For $V\rightarrow\infty$ the coexistence of both phases requires
$P_{\rm h}=P_{\rm q}$, which is the Gibbs condition for mechanical two phase 
equilibrium. This holds also for the case of explicit finite size
modifications of the equation of state, because the correction terms vanish 
by definition in the limit $V\rightarrow\infty$.

\section{The equations of state for infinite matter}
\label{inf_sec}
For illustration let us consider first the case of infinite matter equations
of state, i.~e. without explicit finite size corrections.
The constituents of the low energy density phase are
well-established non-strange hadrons up to masses of 2~GeV.
The system is treated as a mixture of relativistic, non-interacting
Bose--Einstein and Fermi--Dirac gases. As the net-baryon number is zero,
the free energy reads
\begin{equation}
F(T,V)=-\sum_i \frac{g_iV}{6\pi^2} \int_0^\infty
\frac{p^4}{\sqrt{p^2+m^2}}\frac{1}{e^{E_i/T}\pm 1}{\rm d}p \quad ,
\end{equation}
where the $+$ stands for fermions, the $-$ for bosons and $g_i$ denotes the
spin and isospin degeneracy of particle species $i$.
To take into account repulsive interactions, all thermodynamic 
quantities are corrected by the Hagedorn factor
$1/(1+\epsilon/4B)$\cite{Hag80}, where $\epsilon$ is the energy density
of the point particles and $B$ is the bag constant. The bag constant 
is chosen as $B=200\;\rm MeV/fm^3$
for all calculations presented here. 

For the equation of state 
of the high energy density phase (the quark-gluon plasma) we take a simple bag 
model EOS for massless quarks, their antiquarks, 
and gluons in an MIT bag of infinite volume, where the number of quark
flavors is $N_{\rm q}=2$:
\begin{equation}
\label{bageos}
F(T,V)=BV- \pi^2 \left(\frac{7}{60}N_{\rm q}+\frac{8}{45}\right) T^4 V \quad .
\end{equation}
In this case the integration in (\ref{masterquant}) can be carried out
easily, rendering:
\begin{equation}
A(T,V)=
\frac{e^{(-\beta(f_h(T)-f_q(T))V)}[-\beta(f_h(T)-f_q(T))V-1]+1}{
[e^{(-\beta(f_h(T)-f_q(T))V)}-1]
[-\beta(f_h(T)-f_q(T))V]}
(A_h(T)-A_q(T))+A_q(T) \, .
\end{equation}
Note that $f_h(T_{\rm C})=f_q(T_{\rm C})$.
The expectation value of an intensive thermodynamic quantity
like pressure, entropy density or energy density at a temperature
$T\ne T_{\rm C}$ can be calculated with this
formula.

\begin{figure}[b]
\vspace*{-1cm}
\centerline{\psfig{figure=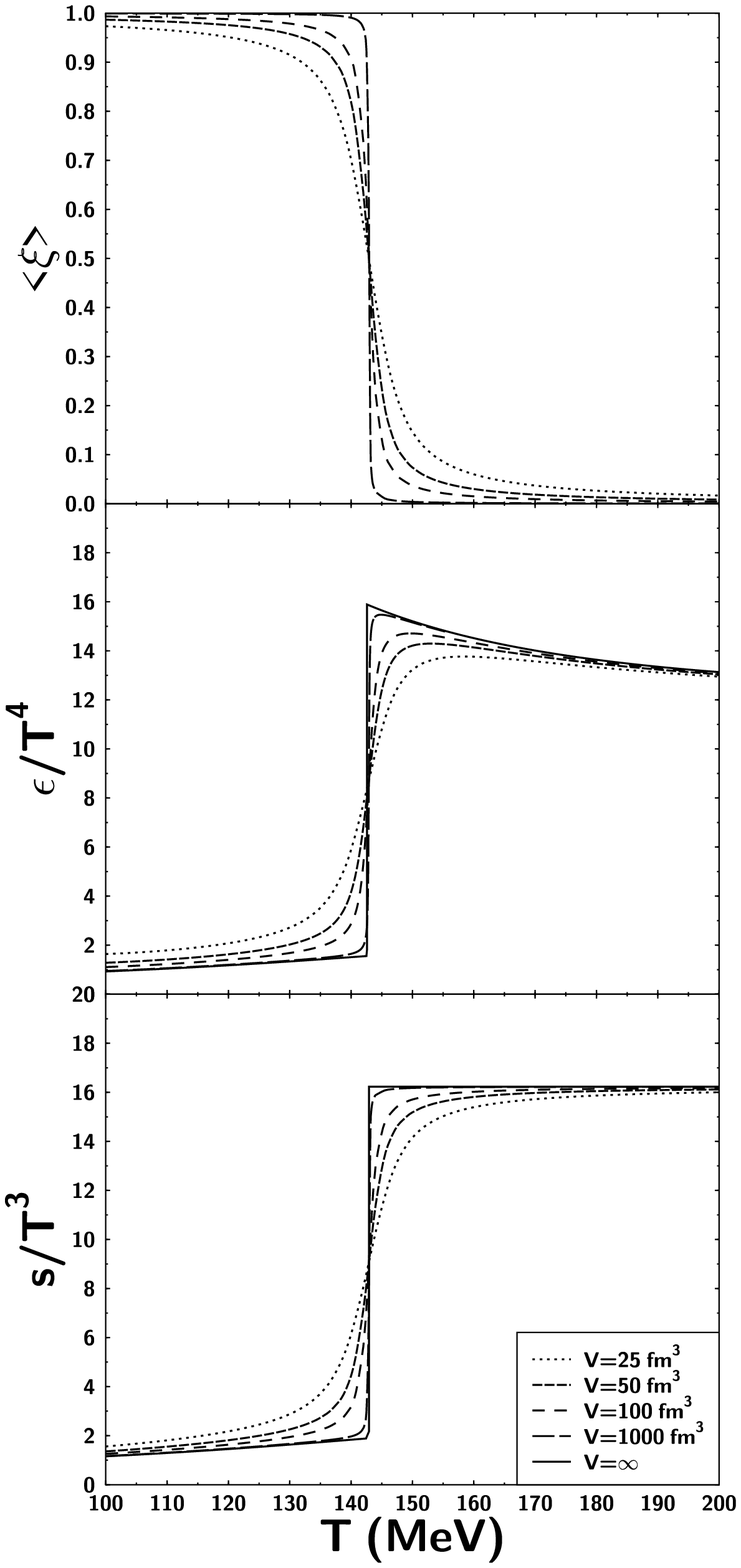,width=12cm}}
\caption{The order parameter (expectation value of the volume fraction of
the hadron gas $<\xi>$) vs.
the temperature for systems of different sizes (top).
The energy density $\epsilon$, divided by the temperature $T$ to the power of 
four (middle),
and the entropy density $s$, divided by the temperature to the power of three
(bottom), are also shown. The bag constant is
$B^{1/4}=200$~MeV. Equations of state for infinite matter (hadronic and QGP)
are used.
\label{3inf}}
\vspace*{-1.5cm}
\end{figure}

Fig.~\ref{3inf} (top) shows the expectation value of the order parameter
$<\xi>$, the average hadronic volume fraction,
as a function of the temperature for different system sizes. This quantity
is calculated as
\begin{equation}
<\xi>(T,V)=\int_0^1{\rm d}\xi\, p(\xi;T,V)\, \xi \quad .
\end{equation}
For volumes $V=100\,\rm fm^3$ or less, $<\xi>(T)$ deviates clearly from a 
simple step function. Even below $T_{\rm C}$, the system is {\em not} purely 
hadronic. The quark phase 
has a finite probability (see also \cite{Sto80}). On the other hand, the hadronic
phase contributes, with similar probability, even above $T_{\rm C}$. 
Note that the order parameter in our
model is simply the volume fraction of the hadronic phase. Thus, for a
volume of $25\,\rm fm^3$ and a temperature of 10~MeV above 
$T_{\rm C}$ one can read off Fig.~\ref{3inf} that 
the system is composed of about 10\ \% hadron gas
and 90\ \% QGP (this admixture for {\em finite} sizes is different from the predicted persistence of
clusters in an {\em infinite} plasma \cite{Ris92}). The unfavorable phase is suppressed,
 according to (\ref{probab}), due to the differences of the free energy.
However, any composition of the total system (any value of $\xi$) must 
correspond to a finite total free energy content, because the volume is finite.
Thus, the suppression relative to the state with minimum free energy content
is finite, i.~e. the probability is non-zero.

Fig.~\ref{3inf} also shows the energy density $\epsilon/T^4$ 
and the entropy density $s/T^3$
 as a function of temperature for different volumes of the
system. In the case of an infinite volume, the first order phase transition
is reflected by the sharp discontinuity of both quantities 
at $T_{\rm C}=143$~MeV. Due 
to the finite probability of fluctuations, the average values of the energy 
density and the entropy density at a temperature $T$ are different for  
finite systems.
Below $T_{\rm C}$, the presence of 
the quark phase, although a small fraction of the total volume, 
increases the energy density and the entropy density. Above
$T_{\rm C}$, the contribution of the hadronic phase still lowers the average
values of $\epsilon$ and $s$ leading to the expected {\em rounding} of the phase transition.
A functional
form of thermodynamic quantities has been 
parametrized \cite{Ris96b,Asa97}; this was used to model the {\em assumed} 
smooth crossover
transition and for studying the resulting physics. 
In any case, for finite size systems such a crossover has to be expected.
Here we calculate such a behaviour for strongly interacting, finite 
systems on the basis of a very simple model without free parameters.

\begin{figure}[b]
\vspace*{\fill}
\centerline{\psfig{figure=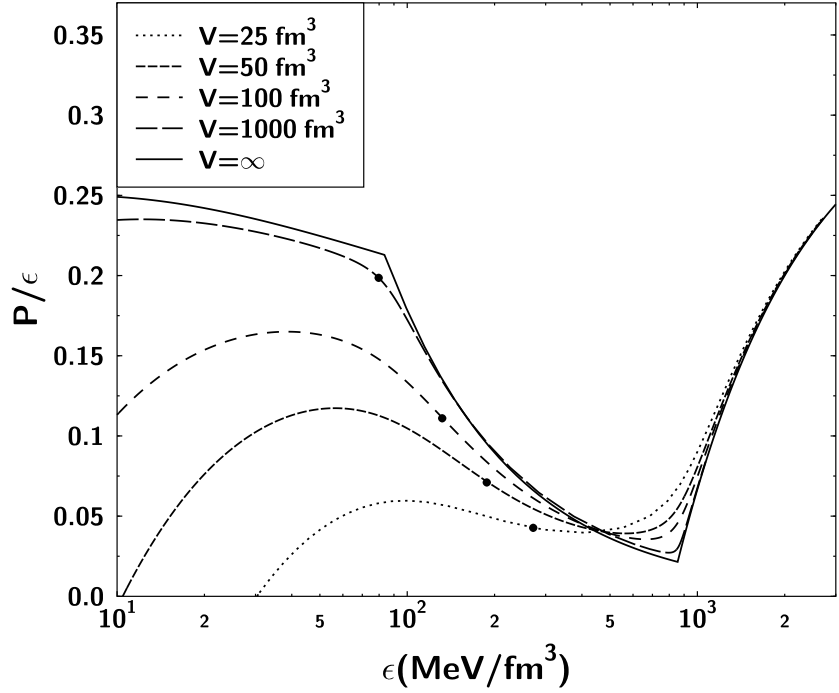,width=12cm}}
\caption{Pressure $P$ divided by the energy density $\epsilon$ 
vs. the energy density for systems of different sizes. The bag constant is
$B^{1/4}=200$~MeV. Equations of state for infinite matter (hadronic and QGP)
are used.
\label{ep}}
\vspace*{\fill}
\end{figure}

Fig.~\ref{ep} now shows the (hydrodynamically relevant) ratio
$P(\epsilon)/\epsilon$ vs.~$\epsilon$ for various system sizes. 
Here the 'softest point' of the 
equation of state vs. $V$ are the respective minima of the curves.
The thermodynamics of the system exhibits a very distinct volume dependence 
with respect to this quantity. Observe the strong influence of $V$ over a wide 
range of energy densities. This could not be clearly seen in
Fig.~\ref{3inf},
where the transition appears within a rather narrow region of temperature.

It has been suggested \cite{Hun95,Ris96a} that a clear peak of the
lifetime of the mixed phase as a function of the collision energy
could signal the QCD phase transition in heavy ion reactions. 
This peak should be observable in a
particular window of the initial energy density, around the 
softest point of the equation of state. The comparatively small pressure
 prevents a fast expansion of the system around that well located point.

However, as can be seen in Fig.~\ref{ep}, the minimum of $P(\epsilon)/\epsilon$ 
is much less pronounced for finite volumes than for an infinite volume. 
At low temperatures --- which
correspond to low energy densities --- the
influence of a small admixture of quark matter on $P$ and
$\epsilon$ is particularly strong, because the absolute values of $P$ 
and $\epsilon$ are
very large in the quark phase in the simple MIT bag description 
($-P=\epsilon=B$ for $T=0$) as compared to the values 
in the hadron phase ($P=\epsilon=0$ for $T=0$). In Fig.~\ref{ep} full
circles on the curves indicate the values of energy density, where the 
pressure of the pure quark phase would become zero according to the
simple bag
model equation of state, eq.~(\ref{bageos}). 
Whether the
concept of thermodynamic fluctuation theory is adequate for lower energy
densities and how the equation of state for the supercooled quark matter
phase might look like at temperatures considerably lower than $T_{\rm C}$
remains to be investigated.

\section{Quark matter equation of state with color singlet constraint}
A grand canonical partition function for a quark-gluon plasma droplet with
the {\em necessary} requirement of color-singletness was proposed in \cite{Elz83}. 
The internal $\rm SU(3)_C$-symmetry is accounted for by applying a 
group-theoretical projection technique \cite{Red80} on color singlet states
to the grand canonical partition function of a noninteracting quantum 
gas of massless quarks and gluons. 
The authors find that, due to the color confinement, the internal degrees of 
freedom in a finite plasma droplet are effectively reduced
as compared to the infinite matter equation of state.
Also, the finite level density (in a finite system) for the single-particle
eigenstates lead to an additional effective reduction. This is explicitely
accounted for.
In the following,
we use the corresponding free energy as the equation of state for the
high density phase. It reads \cite{Elz83}

\begin{equation}
F(T,V)=B\,V - T\,\left( X - Y \right)  + 
  {\frac{3}{2}\,T\,\left( \log (D) - \log (\pi ) \right) } + 
  T\,\left( 4\,\log (C) + \log (2\,{\sqrt{3}}\,\pi ) \right) 
\end{equation}

\[
X=
{\pi^2 \left(\frac{7}{60}N_{\rm q}+\frac{8}{45}\right)
\,{T^3}\,V} \nonumber
\]

\[
Y={{\pi\,\left( \frac{1}{3} N_{\rm q} + \frac{32}{9} 
        \right)\,R \,T}} \nonumber
\]

\[
C=-\frac{1}{\pi} \left(\frac{2}{3} N_{\rm q}-8 \right) \,R\,T + 
  \left(\frac{1}{3} N_{\rm q} +2 \right)
\,{T^3}\,V \nonumber
\]

\[
D=- \pi \left(\frac{1}{9} N_{\rm q} + \frac{32}{27} \right)\,R \,T 
 + \pi^2 \left( \frac{7}{30} N_{\rm q} + \frac{16}{45} \right)
\,{T^3}\,V \quad ,
\]
where $R=(\frac{3V}{4\pi})^{1/3}$ is the radius of the spherical
plasma drop and
$N_q=2$ denotes the number of (massless) quarks.
This equation of state (for the quark-gluon plasma) incorporates
an explicit volume dependence of the pressure at fixed $T$.
In this case, $\displaystyle P_{\rm q}(T)=-\frac{\partial F_{\rm q}}{\partial
V_{\rm q}} \ne -\frac{F_{\rm q}}{V_{\rm q}}$. Therefore, eqs.~(\ref{probab}) and
(\ref{masterquant}) have to be evaluated numerically.
The partition function is derived in saddle-point approximation which
breaks down for $RT\rightarrow 0$. It agrees within $\approx
30$\% with the exact value for $RT\approx 1$\cite{Elz83}. Therefore, we approximate
the free energy density by $f(R)=f(1/T)$ for plasma volumes $(1-\xi)V_{\rm
tot}$, which correspond to spherical droplets of sizes less than $R<1/T$.
The hadronic equation of state of Sec.~\ref{inf_sec} is used, without any
explicit finite size modifications.

\begin{figure}[b]
\vspace*{\fill}
\centerline{\psfig{figure=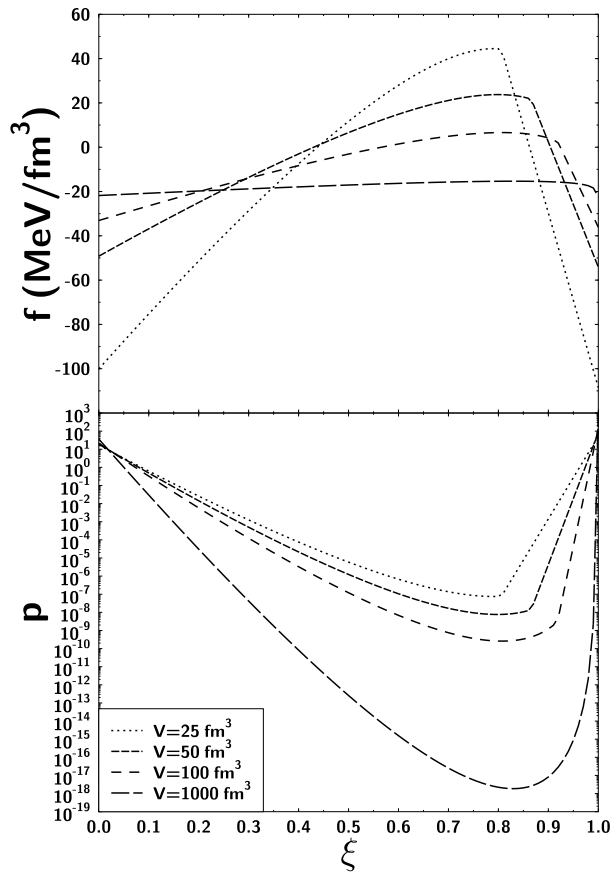,width=12cm}}
\caption{Free energy density $f$ as a function of the hadron fraction for 
systems of different sizes at $T_{\rm C}$ (top) and the resulting 
probability densities $p(\xi)$ (bottom).
The bag constant is
$B^{1/4}=200$~MeV. The color singlet constraint is taken into account for
the QGP equation of state.
\label{2hump}}
\vspace*{\fill}
\end{figure}

Fig.~\ref{2hump} (top) shows the free energy density as a function of the 
order parameter $\xi$ (the hadron volume
fraction) using the color singlet equation of state for the quark phase.
The temperature is chosen to be $T=T_{\rm C}$, where we {\em define} the 
critical temperature at the point where
$\epsilon(T_{\rm C})=(\epsilon_q(T_{\rm C})+\epsilon_h(T_{\rm C}))/2$.
Thus, $T_{\rm C}$ denotes the temperature, for which
both phases contribute with equal probability to the state of the total
system.
In Sec.~\ref{inf_sec} the critical temperature was found to be 
independent of the volume --- only the {\em width} of the transition was 
affected  by the flucutations.

For the finite size corrected equation of state, however, the critical temperature 
shifts to higher values at finite volumes (see below).
As one expects, the finite size modifications of the equation of state
affect the free energy density most strongly for small systems. The infinite
volume limit converges to a constant free energy density.
Fig.~\ref{2hump} (bottom) shows the resulting probability densities as
functions of the order parameter $\xi$. A `two-hump' structure is observed,
thus favoring the dominance of one of the phases against a two phase composition
with equal volume fractions for both phases. 
Such a two-hump structure is generally expected to occur for first order
phase transitions \cite{Bin84}.

One should emphasize that the peaks at $\xi=0$
and $\xi=1$ are most pronounced for the {\em largest} volume. This is
because the free energy, and not the free energy density, enters
eq.~(\ref{probab}) for the calculation of the probability density. The physics
of large systems is therefore extremely sensitive to small variations of the
free energy density.
Because of this very distinct hump structure, either one phase or the other is
present at $T_{\rm C}$ (which 
results from the explicit finite size
corrections). Then, a first order phase transition occurs
 dynamically in a way
that the high temperature phase (or the low temperature phase, respectively)
is supercooled (or superheated, respectively), up to a point where bubble
nucleation starts to convert the now unstable phase to the more stable phase
\cite{Kaj86}.

\begin{figure}[b]
\vspace*{-1cm}
\centerline{\psfig{figure=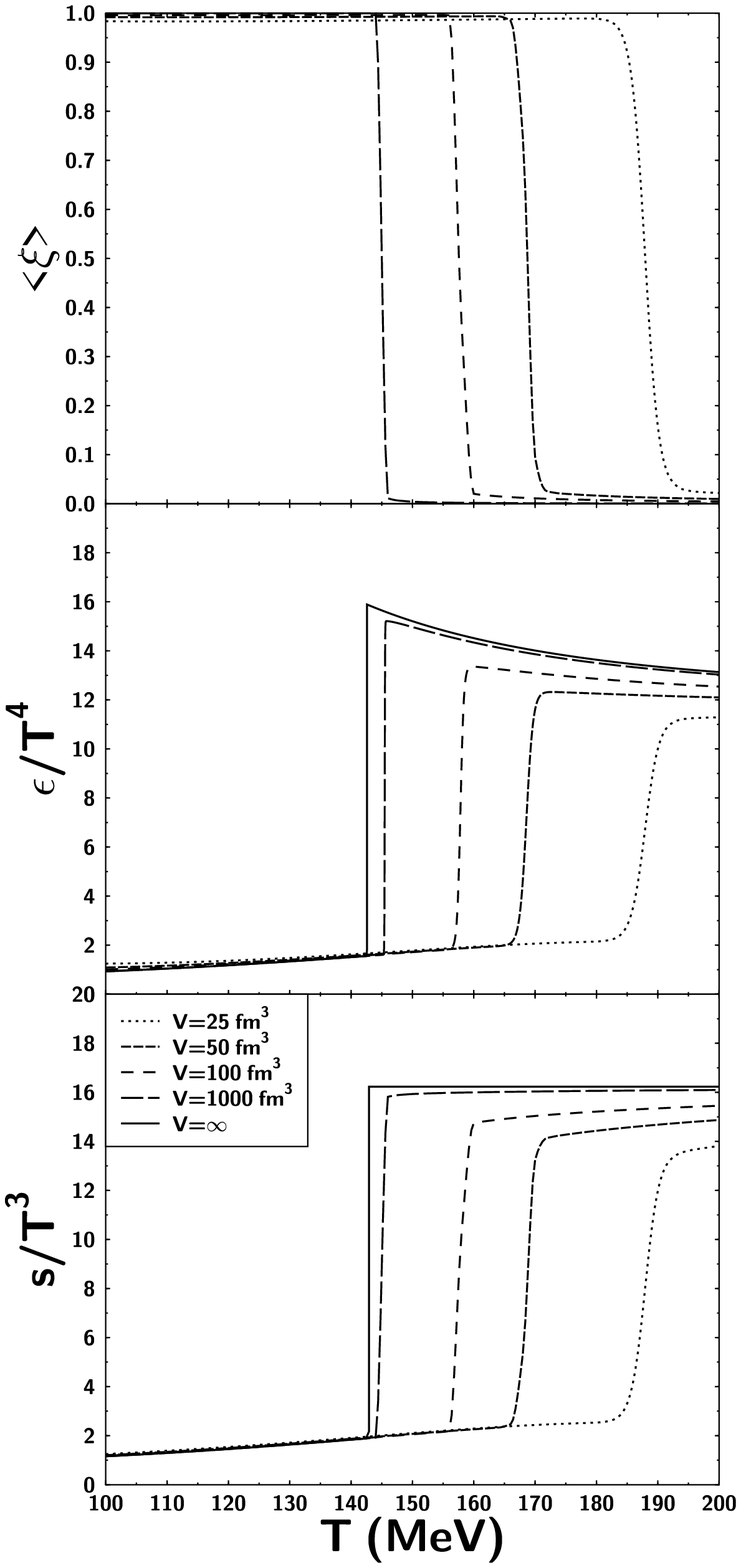,width=12cm}}
\caption{The order parameter (expectation value of the volume fraction of
the hadron gas $<\xi>$) vs.
the temperature for systems of different sizes (top).
The energy density $\epsilon$, divided by the temperature $T$ to the power of 
four (middle),
and the entropy density $s$, divided by the temperature to the power of three
(bottom), are also shown. The bag constant is
$B^{1/4}=200$~MeV. 
The color singlet constraint is taken into account for
the QGP equation of state.
\label{3fin}}
\vspace*{-1.5cm}
\end{figure}

Fig.~\ref{3fin} shows the temperature dependence of the order parameter, the 
energy density  and the entropy density, as in Fig.~\ref{3inf}, but now for
the quark matter equation of state with the color singlet constraint.
The main difference is the shift of the critical temperature to higher
values for smaller volumes.
For systems of finite size the modified equation of state yields a lower
pressure at a given temperature, because the internal degrees of freedom
are gradually 'frozen' with decreasing volume\cite{Elz83}.
Thus, even when flucuations are neglected, the mechanical Gibbs equilibrium
between the two pressures would be reached for temperatures 
$T_{\rm C}>T^{\infty}_{\rm C}$. Here $T^{\infty}_{\rm C}$ stands for the critical
temperature of the {\em infinite} system.

Important consequences follow for the bulk quantities $\epsilon/T^4$ and
$s/T^3$: the latent heat and the jump in the entropy density are considerably 
reduced for small systems. First, the effective number of degrees of freedom
in the hadron phase increases with temperature: the restmasses
become more and more negligible as compared to the kinetic energies. Secondly, as
mentioned before, the effective number of degrees of freedom in the quark phase 
is reduced for finite volumes due to the requirement of color singletness.
As stated in \cite{Asa97}, the active degrees of freedom, which are
``quantified'' as $s(T)$, determine completely the gross behaviour 
of the thermodynamics near the phase transition.
Fig.~\ref{3fin} also shows that the smearing due to the
fluctuations is less pronounced than for the infinite matter 
equation of state in Sec.~\ref{inf_sec}. 
The rounding effect of a fluctuating phase composition
of the system appears to be counteracted by the effect of the volume dependent
reduction of quark-gluon degrees of freedom.

\begin{figure}[b]
\vspace*{\fill}
\centerline{\psfig{figure=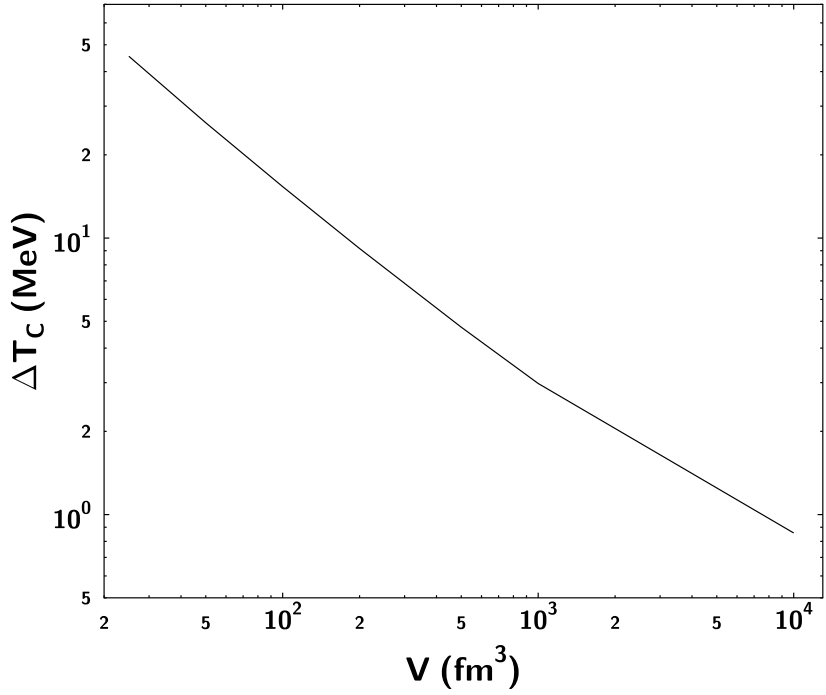,width=12cm}}
\caption{Shift of the critical temperature $\Delta T_{\rm C}$ vs.
the systems size $V$. The bag constant is
$B^{1/4}=200$~MeV. The color singlet constraint is taken into account for
the QGP equation of state. 
\label{deltatc}}
\vspace*{\fill}
\end{figure}

Fig.~\ref{deltatc} depicts the volume dependence of the shift of the
critical temperature, $\Delta T_{\rm C}=T_{\rm
C}-T_{\rm C}^{\infty}$. The temperature shift  
exhibits an approximate power law, $\Delta T_{\rm C} \sim 1/V^\alpha$
with $\alpha \approx 0.7$. The analysis of Binder and Landau
\cite{Bin84}, who examined the first order phase transition of a finite system
in the Ising ferromagnet model, advocates a coefficient $\alpha=1$.

\begin{figure}[b]
\vspace*{\fill}
\centerline{\psfig{figure=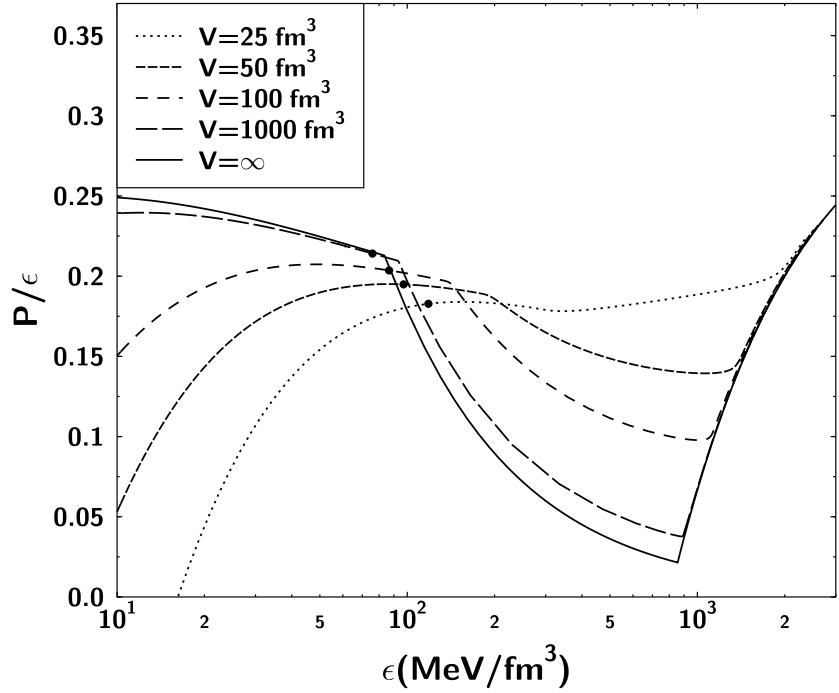,width=12cm}}
\caption{Pressure $P$ divided by the energy density $\epsilon$ 
 vs. the the energy density for systems of different sizes. The bag constant is
$B^{1/4}=200$~MeV. The color singlet constraint is taken into account for
the QGP equation of state.
\label{cep}}
\vspace*{\fill}
\end{figure}

As in Fig.~\ref{ep}, Fig.~\ref{cep} shows the ratio $P(\epsilon)/\epsilon$ 
vs.~$\epsilon$ within the present scenario. Again, 
those values of the energy density are marked, 
which correspond to pressure zero of the pure quark phase.
In clear contrast to Fig.~\ref{ep}, the 'softest point' is now characterized 
by less pronounced minima of
$P(\epsilon)/\epsilon$ for reasonably small system sizes. 
For volumes of $V<25\,\rm
fm^3$, the minimum of the curve vanishes completely. Hence,
the lifetime signal \cite{Hun95,Ris96a},
 which is based on
hydrodynamic considerations and infinite matter equations of state,
will be extremely damped in a more realistic scenario of heavy ion collisions.

\begin{figure}[b]
\vspace*{\fill}
\centerline{\psfig{figure=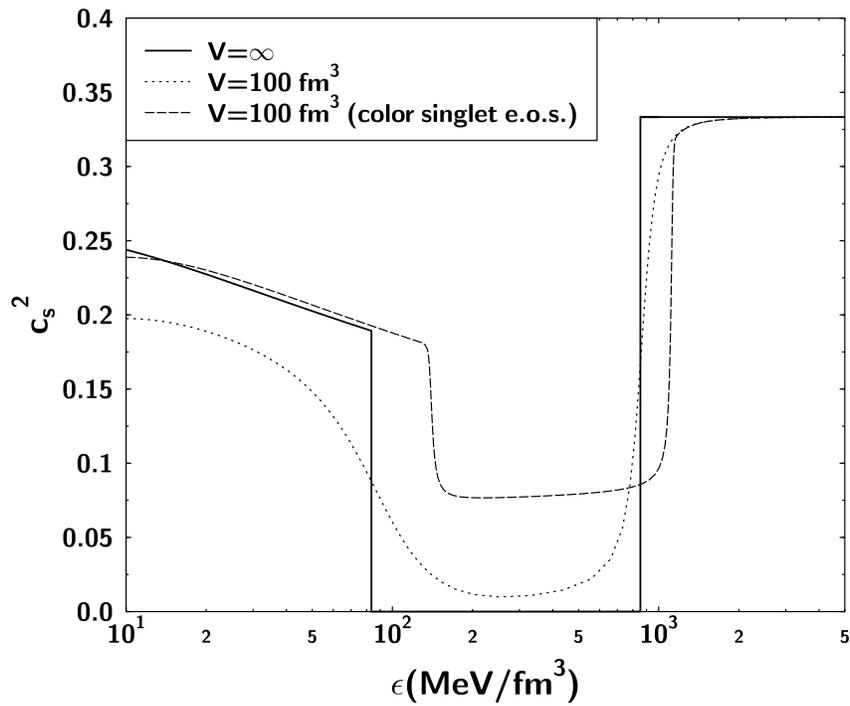,width=12cm}}
\caption{Speed of sound (squared) $c_s^2$ as a function of energy density
$\epsilon$ for three different
cases:
1) infinite volume of the system (full line). \\
2) $V=100\,\rm fm^3$ using the infinite matter EOS (dotted). \\
3) $V=100\,\rm fm^3$ using the EOS with color singlet constraint (dashed).
\label{cs2}}
\vspace*{\fill}
\end{figure}

Fig.~\ref{cs2} shows the speed of 
sound (squared) $\displaystyle c_s^2=\frac{\partial p}{\partial\epsilon}$
 as a function of the energy density for three different
cases. For infinite volume, the phase transition is truely first
order. This can be seen from the vanishing speed of sound in the mixed
phase. This is the cause of the pronounced time delay in hydrodynamical
simulations: even if there are strong gradients in the energy
density, the mixed phase cannot perform mechanical work.
Therefore, it does not expand on its own account. Deflagration fronts with
small velocities convert the mixed phase into hadrons, leading to slow
cooling and expansion \cite{Ris96b}. 

A finite system of volume $V=100\;\rm fm^3$ is also 
depicted in Fig.~\ref{cs2} for the two different equations of
state of the QGP. Neglecting the color singlet constraint (but including the
effect of thermodynamic flucutations in finite volumes) leads to a strong
reduction of the speed of sound in the transition region. Thus, one still
might expect a moderatively long lived fireball of the mixed phase. 
However, the
speed of sound does not vanish in a sharp region of energy density. Rather,
it is smoothly damped in the large $\epsilon$ domain 
between $20\,\rm MeV/fm^3$ and $200\,\rm MeV/fm^3$.
This effect would vastly smear out the lifetime signal if one scans the flow
excitation function over a wide range of collision energies.

After the requirement of color singletness is included, 
$c_s^2$ is much less reduced. The dip is shifted to higher (factor $\sim 2$)
energy densities. Although the rather sharp breakdown of the speed of sound
in a
well located energy density domain is recovered, the hydrodynamical expansion
solutions will now look much different from the infinite matter scenario:
The rather high values of
$c_s^2$ must lead to a much more rapid expansion
than for infinite
matter equations of state. 
The implementation of the
 present results, based
on finite volumes and global equilibrium and also supercooling, into
hydrodynamical
calculations, remains a difficult enterprise for future investigations.

\section{Conclusion}
We have discussed the 
thermodynamic bulk
properties for finite systems at the phase transition and the detailed 
behaviour of the free
energy density as a function of the temperature.
For finite volumes, $V<100\; \rm fm^3$, corresponding to the expected plasma 
volumes, there is a considerable
rounding in the variables $\epsilon/T^4$ and
$s/T^3$ around $T_C$.
As a consequence, the speed of sound does not vanish in the mixed phase.
This is infered, under simple
assumptions, from basic thermodynamic considerations.
Fluctuations of the two phases in a finite
system lead to a smooth transition between the low temperature regime --- where
the hadronic phase dominates the system --- and the high temperature regime
--- where the pure quark phase is most probable. 

The requirement of exact 
color-singletness within the quark phase leads to a shift of the critical
temperature to higher temperatures for finite volumes. We observe a double 
hump structure in the
probability distribution for the actual phase composition during the phase
transition. 
In the model scenario the speed of sound is considerably increased
in the mixed phase, if the requirement of color-singletness is taken into
account. The significance of the time delay signal \cite{Hun95,Ris96a} for
the experimental detection of a QGP phase in heavy ion
becomes questionable. 

In addition, our investigations also show intuitively that the
extrapolation from the behaviour of a finite system to its infinite volume
limit may, in fact, be rather model dependent. 

\acknowledgements
We thank D.~Rischke and A.~Dumitru for fruitful discussions. C.~G. also
would like to thank L.~Csernai and T.~Biro for helpful remarks in the early
stage of the work. This work was supported by the
Gesellschaft f\"ur Schwer\-io\-nen\-for\-schung, Darmstadt, Germany,
the Bundesministerium
f\"ur Forschung und Technologie, Bonn, Deutsche Forschungsgemeinschaft, and
the Graduiertenkolleg Schwerionenphysik (Frankfurt/Giessen).

\begin{appendix}
\section{Thermodynamic derivation of the model}
For a canonical ensemble all thermodynamic quantities like the free energy $F$,
the entropy $S$, the pressure $P$ and the energy $E$ are determined by the
partition function $\cal Z$:
\be
{\cal Z}={\rm Tr}\{e^{-\beta {\hat H}}\}=\sum_n e^{-\beta E_n} \quad , \quad
F=-T\ln{\cal Z} \quad .
\ee
Once $F(T,V)$ is fixed, $S$, $P$, $E$ follow from standard relations.

The main assumption for the case of two coexisting 
phases is {\em separability} of the energy spectra $E^{(i)}_{n_i}$ of the 
single phases $i$. 
It is assumed that
the energy eigenvalues of the total system $E_{\bar n}$ factorize:
\be
\label{eqa6}
E_{\bar n}(\xi)=E^{(1)}_{n_1}(\xi)+E^{(2)}_{n_2}(\xi) \quad .
\ee
Here $\xi$ is a not yet specified order paramter which characterizes the 
configuration
of the total system, i.~e. the relative importance of the individual phases.
The partition function for a given $\xi$ of the two phase system then reads:
\bea
{\cal Z}(\xi)&=&{\rm Tr}_{\bar n}\{e^{-\beta {\hat H}}\}=\sum_{n_1} e^{-\beta
E_{n_1}}\sum_{n_2} e^{-\beta E_{n_2}}={\cal Z}^{1}(\xi){\cal Z}^{2}(\xi)
\quad ,\nonumber \\
F(\xi)&=&-T\ln{\cal Z}(\xi)=F^{(1)}(\xi)+F^{(2)}(\xi) \quad .
\label{eqa8}
\eea
(\ref{eqa6}) guarantees that the simple factorization in (\ref{eqa8}) holds.
Since $\xi$ is a (continuous) order parameter, any value of $\xi$
corresponds to a different state of the two phase system. As in our
intuitive choice for $\xi$ in the main text, we now assume that, 
without loss of generality, $\xi$ can take any number in the range 
between 0 and 1.
The total partition function of the system for a discretized 
spectrum of the order 
parameter and a given total volume $V^{\rm tot}$ reads:
\be
{\cal Z}\equiv {\cal Z}(\xi_1)+{\cal Z}(\xi_2)+\dots {\cal Z}(\xi_n) \quad ,
\mbox{with}\quad
\xi_i=\frac{(i-1)}{n}  \quad .
\ee
The probability for the system being in the state $\xi_i$ is then given by
\be
\Delta p(\xi_i)=\frac{{\cal Z}(\xi_i)}{\cal Z} \quad , \quad \sum_{i=1}^n \Delta
p(\xi_i)\equiv 1  \quad .
\ee
From this we infer the probability density of the continuous case:
\be
p(\xi)=\frac{{\cal Z}(\xi)}{\int_0^1{\cal Z}(\xi){\rm d}\xi} \quad .
\ee
We now evaluate the free energy of the total system 
$F^{\rm tot}$ as well as the other thermodynamic quantities $S^{\rm tot}$,
$P^{\rm tot}$ and $E^{\rm tot}$. In the discretized case we have
\be
\label{eqztot}
{\cal Z}=\sum_{i=1}^n{\cal Z}(\xi_i)=\sum_{i=1}^n e^{-\beta F(\xi_i)}\equiv
e^{-\beta F^{\rm tot}}\quad .
\ee
Differentiating with regard to $\beta$ (as an independent paramter)
yields:
\be
\label{eqftot}
F^{\rm tot}=\sum_{i=1}^n \frac{e^{-\beta F(\xi_i)}}{\sum_{j=1}^n 
e^{-\beta F(\xi_j)}} F(\xi_i) \quad \longrightarrow \quad
\int_0^1 p(\xi)F(\xi){\rm d}\xi \quad .
\ee
The entropy is readily obtained as
\bea
S^{\rm tot}&=&-\left(\frac{\partial F^{\rm tot}}{\partial T}\right)|_V=
\ln{\cal Z}+\frac{T}{\cal
Z}\left(\frac{\partial {\cal Z}}{\partial
T}\right)|_V \nonumber \\
&=&\ln{\cal Z}+\frac{T}{\sum_{j=1}^n {\cal
Z}(\xi_j)}\sum_{i=1}^n \left(\frac{\partial {\cal Z}(\xi_i)}{\partial
T}\right)|_V\nonumber \\
&=&\ln{\cal Z}+\frac{T}{\sum_{j=1}^n {\cal
Z}(\xi_j)}\sum_{i=1}^n \left[e^{-\beta F(\xi_i)}\frac{F(\xi_i)}{T^2}-
\beta e^{-\beta F(\xi_i)} 
\left(\frac{\partial F(\xi_i)}{\partial T}\right)|_V\right]\nonumber \\
&=&\underbrace{\ln{\cal Z}+\frac{1}{T}\sum_{i=1}^n \frac{e^{-\beta
F(\xi_i)}}{\sum_{j=1}^n
{\cal Z}(\xi_j)}F(\xi_i)}_{\equiv 0}
-\sum_{i=1}^n \frac{e^{-\beta F(\xi_i)}}{\sum_{j=1}^n 
{\cal Z}(\xi_j)}\left(\frac{\partial F(\xi_i)}{\partial
T}\right)|_V\nonumber  \\
&\longrightarrow& \int_0^1 p(\xi)
\left(-\frac{\partial F(\xi)}{\partial T}\right)|_V{\rm d}\xi\equiv
\int_0^1 p(\xi)S(\xi){\rm d}\xi \quad ,
\eea
similar for the pressure,
\bea
P^{\rm tot}&=&-\left(\frac{\partial F^{\rm tot}}{V}\right)|_T=\frac{T}{\cal Z}
\left(\frac{\partial {\cal Z}}{V}\right)|_T\nonumber \\
&=&\frac{T}{\sum_{j=1}^n {\cal
Z}(\xi_j)}\sum_{i=1}^n \left(-\beta\frac{\partial F(\xi_i)}{\partial
V}\right)|_T e^{-\beta F(\xi_i)}\nonumber \\
&=&\sum_{i=1}^n \frac{e^{-\beta F(\xi_i)}}{\sum_{j=1}^n 
{\cal Z}(\xi_j)}\left(-\frac{\partial F(\xi_i)}{\partial
V}\right)|_T \nonumber \\
&\longrightarrow& \int_0^1 p(\xi)
\left(-\frac{\partial F(\xi)}{\partial V}\right)|_T{\rm d}\xi\equiv
\int_0^1 p(\xi)P(\xi){\rm d}\xi \quad ,
\eea
and the energy
\bea
\label{eqenergy}
E^{\rm tot}&=&F^{\rm tot}+T^{\rm tot}S^{\rm tot}\nonumber \\
&=&\sum_{i=1}^n \Delta p(\xi_i)\left[F(\xi_i)-T
\left(\frac{\partial F(\xi_i)}{\partial T}\right)|_V\right]\nonumber \\
&\longrightarrow& \int_0^1 p(\xi)
\left[ F(\xi)-T\left(\frac{\partial F(\xi)}{\partial T}\right)|_V\right]
{\rm d}\xi\equiv
\int_0^1 p(\xi)E(\xi){\rm d}\xi \quad .
\eea
(\ref{eqftot})-(\ref{eqenergy}) do correspond to the prescription
eq.~(\ref{masterquant})
stated within the main text.

We now choose the order parameter $\xi$ as the volume fraction of one of
the two phases. The free energy of the system as a function of the order 
paramter thus becomes (confer to eq.~(\ref{freeenergy}))
\be
F(\xi)=\left[f^{(1)}(\xi V^{\rm tot})\xi+f^{(2)}((1-\xi) V^{\rm
tot})(1-\xi)\right] V^{\rm tot}
\quad ,
\ee
where $f^{(i)}$ is the free energy density of phase $i$. Note that the
individual free energy densities can depend on the volume explicitly. For
infinite matter equations of state, however, this is not the case.
We now get for the other quantities:
\be
S(\xi)=-\left(\frac{\partial F(\xi)}{\partial T}\right)|_V=
\left[s^{(1)}(\xi V^{\rm tot})\xi+s^{(2)}((1-\xi) V^{\rm
tot})(1-\xi)\right] V^{\rm tot}
\quad ,
\ee
where $s^{(i)}$ denotes the entropy densities,
\be
P(\xi)=-\left(\frac{\partial F(\xi)}{\partial V}\right)|_T=
\left[P^{(1)}(\xi V^{\rm tot})\xi+P^{(2)}((1-\xi) V^{\rm
tot})(1-\xi)\right]
\quad ,
\ee
and
\be
E(\xi)=F(\xi)+T(\xi)S(\xi)=
\left[e^{(1)}(\xi V^{\rm tot})\xi+e^{(2)}((1-\xi) V^{\rm
tot})(1-\xi)\right] V^{\rm tot}
\quad ,
\ee
where $e^{(i)}$ are the energy densities.

\end{appendix}

\clearpage

\clearpage

\clearpage

\clearpage

\clearpage

\clearpage

\clearpage

\clearpage

\end{document}